# Application of Random Matrix Theory to Biological Networks

Feng Luo[1], Jianxin Zhong[2,3,*], Yunfeng Yang[4], Richard H. Scheuermann[1], Jizhong Zhou[4,*]

[1] *Department of Pathology, U.T. Southwestern Medical Center,
5323 Harry Hines Blvd. Dallas, TX 75390-9072*
[2] *Metals & Ceramics Division, Oak Ridge National Laboratory, Oak Ridge, Tennessee 37831*
[3] *Department of Physics, Xiangtan University, Hunan 411105, China*
[4] *Environmental Sciences Division, Oak Ridge National Laboratory, Oak Ridge, Tennessee 37831*

We show that spectral fluctuation of interaction matrices of yeast a core protein interaction network and a metabolic network follows the description of the Gaussian orthogonal ensemble (GOE) of random matrix theory (RMT). Furthermore, we demonstrate that while the global biological networks evaluated belong to GOE, removal of interactions between constituents transitions the networks to systems of isolated modules described by the Poisson statistics of RMT. Our results indicate that although biological networks are very different from other complex systems at the molecular level, they display the same statistical properties at large scale. The transition point provides a new objective approach for the identification of functional modules.

The cell is a complex system that contains numerous functionally diverse elements, including protein, DNA, RNA and small molecules. Understanding the fundamental principles and behavioral properties of the cell as a system has become a key research activity in the post-genomic era. Research on the topological properties of large scale networks of cell constituents has shown that biological networks share some fundamental topological properties, including scale-free, small-world, hierarchical, modular [1] and self-similar [2] properties, with other complex systems, such as the internet and social networks. Inspired by the electrical engineering paradigm, small gene circuit descriptions combined with mathematical modeling have been utilized to understand small subsystems of cellular processes [3]. Unfortunately, the huge number of constituents and their complex relationships in the cell make the mathematical modeling of large-scale biological systems challenging. It is of significant importance to understand the nature of the structure and interactions of biological networks for achieving quantitative description of their functions.

In this Letter, we use RMT to analyze the structure and interactions of biological networks. RMT, initially proposed by Wigner and Dyson in the 1960s for studying the spectrum of complex nuclei [4], is a powerful approach for the identification and modeling of phase transitions and dynamics in physical systems. It has been successfully used to study the behaviors of complex systems, such as spectral properties of large atoms [5], metal insulator transitions in disordered systems [6], spectra of quasiperiodic systems [7, 8], chaotic systems [9], brain responses [10], and the stock market [11]. One of the essential statistical properties in the RMT is eigenvalue fluctuation. For real and symmetrical random matrices that represent the time-reversal invariant complex systems, the eigenvalue fluctuations follow two universal laws depending on the correlation property of eigenvalues. Strong correlation of eigenvalues leads to eigenvalue fluctuations described by the GOE. On the other hand, eigenvalue fluctuations follow Poisson statistics if there is no correlation between eigenvalues.

In this study we have found that the spectral fluctuation of a yeast protein-protein interaction network and a yeast metabolic network is described by the GOE statistics. Furthermore, we demonstrate that while each of these global networks belong to the GOE, removal of interactions between constituents identifies a transitions in which the spectral fluctuation approximates the Poisson statistics of RMT resulting in a decoupled network composed of isolated modules. Such a sharp transition provides a new objective approach for the identification of functional modules within global biological networks.

We used the standard spectral unfolding technique in our study. In general, the density of eigenvalues of a matrix varies with its eigenvalue $E_i$ ($i = 1, 2, 3, ... N$), where $N$ is the order of the matrix. In order to observe the universal eigenvalue fluctuations of different matrices, random matrix theory requires spectral unfolding to have a constant density of eigenvalues. To fulfill this, one can replace $E_i$ by the unfolded spectrum $e_i$, where $e_i = N_{av}(E_i)$ and $N_{av}$ is the smoothed integrated density of eigenvalues obtained by fitting the original integrated density to a cubic spline or by local density average. With the unfolded eigenvalues, We calculated the nearest neighbor spacing distribution (NNSD) of eigenvalues, $P(s)$, which is defined as the probability density of unfolded eigenvalue spacing $s = e_{i+1} - e_i$. We know from RMT that $P(s)$ for the GOE statistics closely follows the Wigner-Dyson distribution

$$P_{GOE}(s) \approx \frac{1}{2}\pi s \exp\left(-\pi s^2/4\right).$$

In the case of Poisson statistics, $P(s)$ is given by the Poisson distribution

$$P_{\text{Poisson}}(s) = \exp(-s).$$

The difference between the Wigner-Dyson and Poisson distributions is their behavior at small values of s, where:

$$P_{\text{GOE}}(s \to 0) = 0 \text{ and } P_{\text{Poisson}}(s \to 0) = 1.$$

We applied the random matrix theory to two biological networks of yeast. The first network is the core protein interaction network of yeast obtained from the DIP [12] database (version ScereCR20041003) generated from the filtering of large high-throughput protein interaction data using two different computational methods [13]. After removal of all self-connecting links, the final protein interaction network includes 2609 yeast proteins and 6355 interactions. The second network is the yeast metabolic network constructed by Jeong et al [14] from the data in the WIT database [15]. After removal of redundant links, the final metabolic network has 1511 chemical substrate and intermediate states and 3807 interactions. The original metabolic network is a directed network. To make the metabolism network symmetric for RMT study, we changed the directed network to an undirected network by replacing all directed links in the network with undirected links.

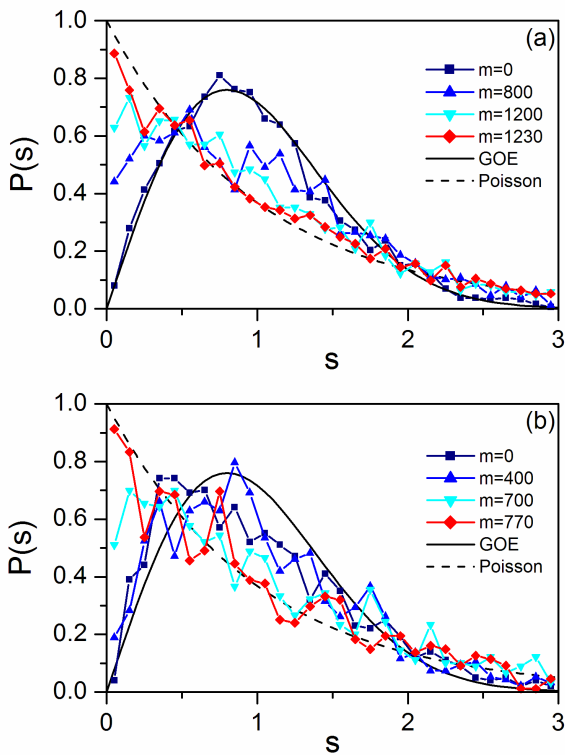

FIG. 1. The NNSDs of yeast biological networks. Smooth and dashed black lines are the GOE distribution and the Poisson distribution, respectively. (a) Yeast core protein-protein interaction networks with different number of removed links (m): 0 (navy), 800 (blue), 1200 (cyan), and 1230 (red). (b) Metabolic networks with different number of removed links (m): 0 (navy), 400 (blue), 700 (cyan), and 770 (red).

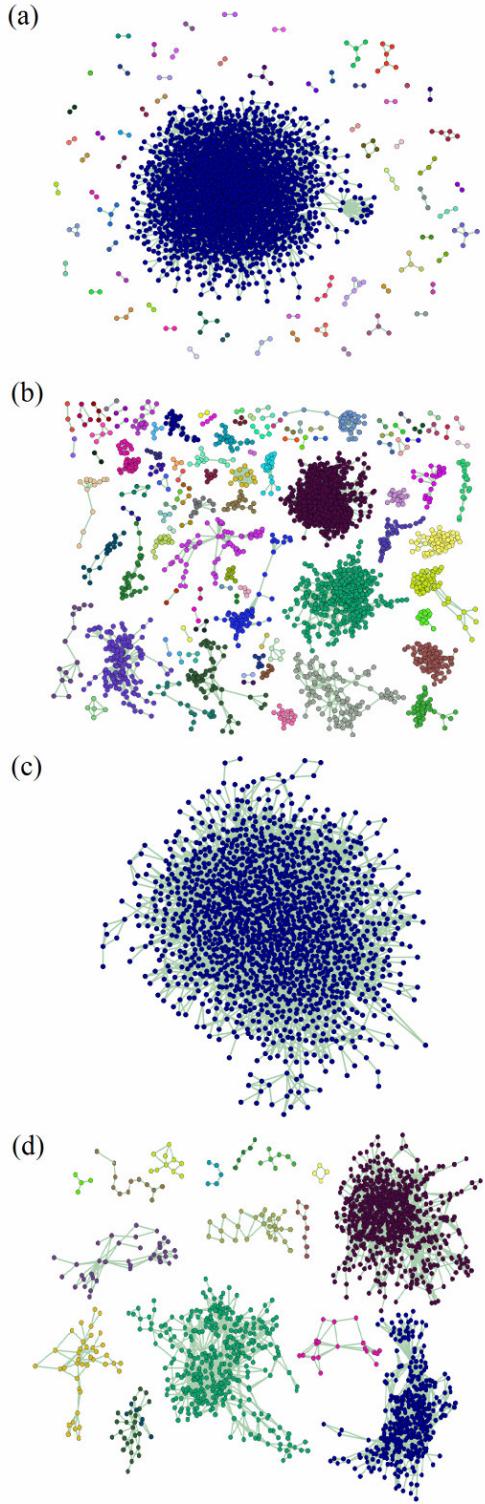

FIG. 2. Graph view of the yeast biological networks. (a) The original yeast core protein interaction network. (b) The yeast core protein interaction network with 1230 links removed. (c) The original yeast metabolic network. (d) The yeast metabolic network with 770 links removed. The graphs were produced using Biolayout [20].

In our RMT analysis, the two biological networks are represented by two real symmetric matrices. The dimension of each matrix is the number of constituents in the network. The elements in the matrices are set to 1 if there is a direct interaction between the constituents; otherwise, the elements are set to 0. We calculated the NNSD of these two matrices for RMT analysis by direct diagonalization of the matrix. Figure 1 shows the NNSDs of these two networks. One can see that the NNSDs of the protein interaction network are well described by the Wigner-Dyson distribution. The NNSDs of the metabolism network are also very close to the Wigner-Dyson distribution, especially in the region representing small values of s. The slight deviation may be due to the incomplete nature of the defined network.

It has been conjectured that biological networks have modular structures with stronger interactions between elements inside the same module and weaker interactions between different modules [16-18]. To test the modularity of these two networks, we gradually removed the interaction links between the constituents in the two yeast networks using the Girvan-Newman algorithm [19] and calculated the NNSDs of the remaining networks. A transition of NNSD from a Wiger-Dyson distribution to a Poisson distribution was clearly observed in both cases (Figure 1). We used the Chi-squared test to determine the transition point. For the DIP yeast core protein interaction network, Chi-squared testing showed that NNSD follows a Poisson distribution after removal of 1230 links. The remaining protein-protein interaction network contains 107 modules with sizes ranging from 2 to 778 proteins. For the yeast metabolic network, NNSD follows a Poisson distribution after removal of 770 links and the remaining network has 17 modules with sizes ranging from 4 to 602 chemical substrates and proteins.

These biological networks can be easily transformed to graphs by representing each element in the network as a vertex and each link as an edge in the graph. Figs. 2(a) and (c) show graph views of the original DIP yeast core protein interaction network and the yeast metabolic network, respectively. Figs. 2(b) and 2(d) illustrate the corresponding networks after removal of links at the transition point. One can see from Fig. 2 that the networks described by the Poisson distribution are very different from the original networks described by the GOE statistics. Isolated modules can be easily identified in Figs. 2(b) and 2(d).

To summarize, we have provided evidence that global biological networks, as represented by the yeast protein-protein interaction and metabolic networks studied here, belong to the GOE. However, by successive removal of interactions between constituents of the network, a global biological network transitions into a system of isolated modules described the Poisson statistic. The transition from a GOE statistic to a Poisson statistic may open a new avenue for objective identification of functional modules inside global networks [21].

ACKNOWLEDGMENTS: This research was supported by The United States Department of Energy under the Genomics: GTL, Microbial Genome Program and Natural and Accelerated Bioremediation Research Programs of the Office of Biological and Environmental Research, Office of Science, and by the National Institutes of Health contracts N01-AI40076 and N01-AI40041. Oak Ridge National Laboratory is managed by University of Tennessee-Battelle LLC for the Department of Energy under contract DE-AC05-00OR22725. We thank Dr. Albert-Laszlo Barabasi, Dr. Hawoong Jeong and Dr. Natalia Maltsev for their help on construction of the metabolic networks.

*Corresponding author.
Electronic address: zhongjn@ornl.gov; zhouj@ornl.gov

[1] Albert-Laszlo Barabasi, Zoltan N. Oltvai, Nature Review, 5, 101, (2004).
[2] C. M. Song, S. Havlin, H. A Makse, Nature, 433, 392, (2005).
[3] J. Hasty, D. McMillen, J. J. Collins, Nature, 420, 14, 224, (2002).
[4] E. P. Wigner, *SIAM Review* 9, 1, (1967).
[5] E. P. Wigner, Proc. Camb. Phil. Soc., 299, 189, (1951).
[6] E. Hofstetter, M. Schreiber, Physical Review. B., 48, 16979 (1993).
[7] J. X. Zhong, U. Grimm, R. A. Romer, M. Schreiber, Physical Review Letter, 80, 3996 (1998).
[8] J. X. Zhong, T. Geisel, *Physical Review* E 59, 4071, (1999).
[9] O. Bohigas, M. J. Giannoni, C. Schmit, *Physical Review Letter* 52, 1 (1984).
[10] P. Seba, Physical Review Letters, 91, 19, 198104, (2003).
[11] V. Plearou, P. Gopikrishnan, B. Rosenow, L. A. N. Amaral, H. E. Stanley, Physical Review Letters, 83, 1471, (1999).
[12] I. Xenarios, L. Salwίnski, X. Q. Duan, P. Higney, S. M. Kim, D. Eisenberg, Nucleic Acids Research, 30, 1, 303, (2002).
[13] C. M. Deane, L. Salwinski, I. Xenarios, D. Eisenberg, Molecular & Cellular Proteomics, 1, 349-356, (2002).
[14] H. Jeong, B. Tombor, R. Albert, Z. N. Oltvai, A.-L. Barabasi, Nature, 407, 651, (2000).
[15] R. Overbeek, et al., Nucleic Acids Res. 28, 123, (2000).
[16] L. H. Hartwell, J. J. Hopfiled, S. Leibler, A. W. Murray, Nature, 402, C47, (1999).
[17] E. Ravasz, A. L. Somera, D. A. Mongru, Z. N. Oltvai, A.-L. Barabasi, Science, 297, 1551, (2002).
[18] A. W. Rives, T. Galitski, PNAS, 100, 1128, (2003).
[19] M. Girvan, M. E. J. Newman, PNAS, 99, 7821, (2002).
[20] A. J. Enright, C. A. Ouzounis, Bioinformatics, 17, 853, (2001).
[21] Feng Luo, Yunfeng Yang, Jianxin Zhong, Jizhong Zhou. Submitted to Nature Biotechnology (2005).